\documentclass[12pt]{article}
\usepackage{graphicx}
\usepackage{amssymb}
\usepackage{latexsym}
\usepackage[dvips]{epsfig}

\newcommand{\pdrv}[2]{ \frac{\partial #1}{\partial #2}}

\newcommand{\pa}{\partial}

\newcommand{\be}{\begin{equation}}
\newcommand{\ee}{\end{equation}}
\newcommand{\bea}{\begin{eqnarray*}}
\newcommand{\eea}{\end{eqnarray*}}
\newcommand{\bean}{\begin{eqnarray}}
\newcommand{\eean}{\end{eqnarray}}
\newcommand{\overleftrightarrow}[1]{\vbox{\ialign{##\crcr
    $\leftrightarrow$\crcr\noalign{\kern-1pt\nointerlineskip}
    $\hfil\displaystyle{#1}\hfil$\crcr}}}
\newcommand{\up}[3]{\stackrel{#1}{#2}\hspace{-3pt}^{#3}}

\newcommand{\td}[1]{\tilde{#1}}

\newcommand{\n}[1]{\label{#1}}

\begin{document}

\title{Scattering of a Long Cosmic String by a Rotating Black Hole}
\author{Martin Snajdr\footnote{e-mail: msnajdr@phys.ualberta.ca} ${}^1$, 
Valeri Frolov\footnote{e-mail: frolov@phys.ualberta.ca} ${}^1$ and 
Jean-Pierre De Villiers\footnote{e-mail: jd5v@vanessa.astro.virginia.edu} ${}^2$}
\maketitle

\begin{center}
\noindent{
$^{1}${\em
Theoretical Physics Institute, Department of Physics, \ University of
Alberta, \\ Edmonton, Canada T6G 2J1}
}

\noindent  {
$^{2}${\em
Department of Astronomy, University of Virginia,
P.O.Box 3818,\\ University Station, Charlottesville, VA 22903-0818}
}
\end{center}
\bigskip

\maketitle

\begin{abstract}   
\noindent  
The scattering of a straight, infinitely long string by a rotating
black hole is considered. We assume that a string is moving with
velocity $v$ and that initially the  string is parallel to the axis
of rotation of the black hole. We demonstrate that as a result of
scattering, the string is displaced  in the direction perpendicular
to the velocity by an amount  $\kappa(v,b)$, where $b$ is the
impact parameter. The late-time solution is represented by a kink and
anti-kink, propagating in opposite directions at the speed of light,
and leaving behind them the string in a new ``phase''.  We present
the results of the numerical study of the string scattering and their
comparison with  the weak-field approximation, valid where the impact
parameter  is large, $b/M \gg 1$, and also with the scattering by a 
non-rotating black hole which was studied in earlier works.  
\end{abstract}

\vspace{3cm}

\newpage
\section{Introduction}
\setcounter{equation}0

Study of cosmic strings and other topological defects and their
motion in an external gravitational field is an interesting problem.
Cosmic strings are topologically stable one-dimensional objects
which are predicted by unified theories. Cosmic strings (as well as
other topological defects) may appear during a phase  transition in
the early Universe. A detailed discussion of cosmic  strings and
other topological defects can be found in the book by Shellard and
Vilenkin \cite{ShVi:94}. Cosmic strings are naturally predicted by
many realistic models of particle physics. The formation of cosmic
strings helps to successfully exit the inflationary era in a number
of the inflation models motivated by particle physics
\cite{KoLi:87,LiRi:97}. The formation of cosmic strings is also
predicted in the most classes of superstring compactification
involving the spontaneous breaking of a pseudo-anomalous $U(1)$ gauge
symmetry ( see e.g. \cite{BiDePe:98} and references therein).
Recent measurements of CMB anisotropy and especially the position of the
acoustic peaks exclude some of the earlier proposed scenarios where
the cosmic strings are the main origin of CMB fluctuations. On the
other hand, recent analysis shows that a mixture of inflation and
topological defects is consistent with current CMB data
\cite{PoVa:99,Cant:00,SiPe:01,LaSh:02,BoPeRiSa:02}.

A black hole interacting with a
cosmic string is a quite rear example of interaction of two
relativistic non-local gravitating systems which allows rather
complete analysis. This makes this system interesting from pure
theoretical point of view. From more "pragmatic" viewpoint, this
system might be a strong source of gravitational waves. But in order
to be able to study this effect one needs first to obtain an information
on the motion of the string in the black hole background. The
situation here is very similar to the case of gravitational radiation
from bodies falling into the black hole.

In this paper we consider the scattering of a long cosmic string by a
rotating black hole. 
We assume that a string is thin and its
gravitational back reaction can be neglected. If $\eta$ is a
characteristic energy scale of a phase transition responsible for a
string formation then the thickness of a string is $\rho \sim
\eta^{-1}$ while its dimensionless mass per  unit length parameter
$\mu^*=G\mu/c^2 \sim \eta^{2}$. For example, for GUT strings
$\rho_{GUT}\approx 10^{-29}$cm and $\mu^*_{GUT}\approx 10^{-6}$, 
and for the electroweak phase transitions 
$\rho_{EW}\approx 10^{-14}$cm and $\mu^*_{EW}\approx 10^{-34}$.
Since $\mu^*\ll 1$ one can neglect (at least in the lowest order
approximation) effects connected with gravitational waves radiation
during the scattering of the string by a black hole. 

We assume that the size of the string is much larger then the
Schwarzschild gravitational radius $r_S=2GM/c^2$ of a black hole and
its total mass is much smaller than the black hole mass. The latter
condition together with $\mu^*\ll 1$ means that we can consider a
string as a test object. In order to specify the scattering problem we
consider the simplest setup, with the string initially far
away from the black hole, so that the string has the form of a straight line.
 We assume that
string is initially moving with the velocity ${\bf v}$ towards the
black hole and lying in the plane parallel to the rotation axis of 
the black hole.
This plane is located at distance $b$ with respect to the parallel ``plane''
passing through the rotation axis.
In analogy with a particle
scattering we call $b$ the {\em impact parameter}.

The scattering of a cosmic string by a {\em non-rotating} black hole was
previously studied both numerically and analytically
\cite{DVFr:97,DVFr:98,Page:98,DVFr:99,Page:99}.
 For $b\gg r_S$ the
string moves in the region where the gravitational potential $GM/r$ is
always small, and one can use a weak-field approximation where the string
equation of motion can be solved analytically \cite{DVFr:98,Page:98,Page:99}.

String scattering in the weak field limit can be described in qualitative terms.
In the reference frame of the string the gravitational field
of the black hole is time dependent and it excites the string's
transversal degrees of freedom. This effect occurs mainly when the
central part of the string passes close to the black hole. Since
information is propagating  along the string at the velocity of
light, there always exist two distant regions, `right' and `left',
which  have not yet felt this excitation. This asymptotic regions of the
string continue their motion in the initial plane (`old phase').  After
scattering, when string is moving again far from the black hole,  its
central part moves again in a plane which is parallel to the
initial plane but is shifted below it by a distance
$\kappa=2\pi GM v/\sqrt{c^2-v^2}$ (`new
phase'). There exist two symmetric kink-like regions separating the
`new' phase from the `old' one. These kinks move out of the central region
with the velocity of light and preserve their form. Besides the amplitude
$\kappa$, the kinks are characterized by a width $w$ which depends on
the impact parameter and the initial velocity of the string.

In the strong-field limit this picture qualitatively
remains the same for string scattering by a
Schwarzschild black hole. This problem was
analyzed in detail  numerically in  \cite{DVFr:98}. In particular, the
quantities $\kappa$ and $w$ were found for different values of $b$
and $v$. The numerical calculations in the strong field limit
demonstrated also that for special values of the initial impact
parameter $b<b_{crit}(v)$ a cosmic string is captured by the black
hole. The critical impact parameter for capture by the
Schwarzschild black hole was obtained in \cite{DVFr:97,DVFr:99}.

In this paper we study of scattering of straight cosmic strings by a
{\em rotating} black hole. We assume that the initial direction of the
string is parallel to the rotational axis of the
black hole. Our focus is on the calculation of the same
quantities, $\kappa$ and  $w$, and to study
how black hole rotation modifies the earlier results for a
non-rotating black hole. We present here results for string
spacetime evolution, real profiles of a string at different external
observer times $T$, as well as asymptotic scattering data. Results
connected with capture of the string and near-critical scattering will
be discussed in another paper.

The paper is organized as follows. Equations for string motion in a
curved spacetime  and their solutions in the weak field
regime are discussed in section~2. Section ~3 contains the
formulation of the initial and boundary conditions for straight
string motion in the Kerr geometry, and the general scheme of the numerical
calculations. In section~4 we describe results for strong field
scattering of strings in the Kerr spacetime and ``real time'' profiles
of the strings. Section~5 contains information about asymptotic
scattering data (displacement parameter and kink profiles) and their
dependence on the impact parameter, velocity, and angular velocity of
the black hole. Section~6 contains a general discussion. Numerical
details are supplied in an appendix.

\section{Equations of motion}
\setcounter{equation}0

\subsection{String equation of motion}

The worldsheet swept by the string in (4D curved background)
space-time can be parametrized by a pair of variables $\zeta^A$
($A=0,1$) so that string motion is described by the equation
$x^{\mu}={\cal X}^{\mu}(\zeta^A)$. The dynamics of the string in a
spacetime with metric $g_{\mu\nu}$ is
described by the Nambu-Goto effective action
\be\n{2.1}
I_0[{\cal X}^{\mu}] = -\mu \int d^2\zeta\ \sqrt{-\gamma}\ ,
\ee 
where 
\be\n{2.2}
\gamma_{AB}=g_{\mu\nu}{\cal X}^\mu_{,A}\, {\cal X}^\nu_{,B}
\ee 
is the induced metric on the worldsheet.
The same equations of motion of the string can be derived from  the
Polyakov's form of the action  \cite{Poly:81},
\be\n{2.3} 
I[{\cal X}^{\mu},h_{AB}]=-{\mu \over 2}\,\int d^2\zeta
\sqrt{-h}h^{AB}\gamma_{AB}\, .
\ee
We use units in which $G=c=1$, and the sign conventions of \cite{MTW}. In 
(\ref{2.3}) $h_{AB}$ is the internal metric
with determinant $h$.

The variation of the action (\ref{2.3}) with respect to ${\cal X}^{\mu}$
and $h_{AB}$ gives the following equations of motion:
\be\n{2.5} 
\Box {\cal X}^{\mu}+h^{AB}{\Gamma^{\mu}}_{\alpha\beta}{\cal
X}^{\alpha}_{,A}{\cal X}^{\beta}_{,B}=0\, ,
\ee
\be\n{2.6} 
\gamma_{AB}-{1\over 2}h_{AB}h^{CD}\gamma_{CD}=0 \, ,
\ee
where
\be\n{2.7} 
\Box ={1\over \sqrt{-h}}\partial_A(\sqrt{-h}h^{AB}\partial_B)\, .
\ee
The first of these equations is the dynamical equation for string
motion, while the second one plays the role of a constraint. 

We fix internal coordinate freedom by using the gauge in which
   $h_{AB}$ is conformal to the flat two-dimensional metric
   $\eta_{AB}=\mbox{diag}(-1,1)$.
In this gauge
the equations of motion for the string have the form 
\be\n{2.8}
 \Box {\cal X}^\mu + {\Gamma^\mu}_{\alpha\beta}\, {\cal
 X}^{\alpha}_{,A}\, {\cal X}^{\beta}_{,B}\, \eta^{AB}\, =0\, ,
\ee 
and the constraint equations are
\be\n{2.9}
\gamma_{01} = g_{\mu\nu}\pdrv{{\cal X}^\mu}{\tau}\pdrv{{\cal X}^\nu}{\sigma} = 0\ 
,
\ee
\be\n{2.10}
\gamma_{00}+\gamma_{11} = g_{\mu\nu}\left(\pdrv{{\cal X}^\mu}{\tau}
\pdrv{{\cal X}^\nu}{\tau}+\pdrv{{\cal X}^\mu}{\sigma}\pdrv{{\cal X}^\nu}{\sigma}
\right)\ = 0\ .
\ee
Here, $\tau\equiv\zeta^0$, $\sigma\equiv\zeta^1$ and 
$\Box=-\pa^2_\tau+\pa^2_\sigma$.
These constraints
are to be satisfied for the initial data. As a consequence of the
dynamical equations they are valid for any later time.

\subsection{Weak field approximation}

In the absence of the external gravitational field
$g_{\mu\nu}=\eta_{\mu\nu}$, where $\eta_{\mu\nu}$ is the flat spacetime
metric. In Cartesian coordinates ($T,X,Y,Z$),
$\eta_{\mu\nu}=\mbox{diag}(-1,1,1,1)$ and
${\Gamma^{\mu}}_{\alpha\beta}=0$, and it is easy to verify that
\be\n{2.11} 
{\cal X}^{\mu}={\cal X}_0^{\mu}(\tau,\sigma)\equiv (\tau\cosh\beta,
\tau\sinh\beta+X_0, b,\sigma) \, ,
\ee
\be\n{2.12} 
h_{AB}=\eta_{AB}\equiv \mbox{diag}(-1,1)\, ,
\ee
is a solution of equations (\ref{2.5}) and (\ref{2.6}). This
solution describes a straight string oriented along the $Z$-axis
which moves in the $X$-direction with constant velocity $v=\tanh
\beta$. Initially, at ${\tau}_{0} = 0$, the string is found at 
${\cal X}^{\mu}(0,\sigma) = (0,X_0, b,\sigma)$. Later when we use
this solution as initial data for a string scattering by a black
hole,   $b$  will play the role of impact parameter. For definiteness
we choose $b>0$ and $X_0<0$, so that $\beta>0$.

Let us consider how this solution is modified when the straight string
is moving in a weak gravitational field. We assume 
\be\n{2.13}
g_{\mu\nu}=\eta_{\mu\nu}+q_{\mu\nu}\, ,
\ee
\be\n{2.14}
{\cal X}^{\mu}(\zeta)={\cal X}_0^{\mu}(\zeta)+\chi^{\mu}(\zeta)\, ,
\ee
where $q_{\mu\nu}$ is the metric perturbation and $\chi^{\mu}$ is the
string perturbation. By making the perturbation of the equation of
motion one gets
\be\n{2.15}
\Box \chi^{\mu}=f^{\mu}\, ,
\ee
where
\be\n{2.16}
f^{\mu}=f^{\mu}(\zeta)=-{\Gamma^{\mu}}_{\alpha\beta}({\cal X}_{0})\, {\cal
X}_{0,A}^{\alpha}\, {\cal X}_{0,B}^{\beta}\, \eta^{AB}\, .
\ee
We use Cartesian coordinates here so that ${\Gamma^{\mu}}_{\alpha\beta}$
is simply the Christoffel symbol for $q_{\mu\nu}$.

The linearized constraint equations are
\be\n{2.17}
\eta_{\mu\nu}\pdrv{{\cal X}^\mu_0}{\tau}\pdrv{\chi^\nu}{\sigma}+
\eta_{\mu\nu}\pdrv{\chi^\mu}{\tau}\pdrv{{\cal X}^\nu_0}{\sigma} +
q_{\mu\nu}\pdrv{{\cal X}^\mu_0}{\tau}\pdrv{{\cal X}^\nu_0}{\sigma} = 0\ ,
\ee
\be\n{2.18}
2\eta_{\mu\nu}\left(\pdrv{{\cal X}^\mu_0}{\tau}\pdrv{\chi^\nu}{\tau}+
\pdrv{{\cal X}^\mu_0}{\sigma}\pdrv{\chi^\nu}{\sigma} \right)+
q_{\mu\nu}\left(\pdrv{{\cal X}^\mu_0}{\tau}\pdrv{{\cal X}^\nu_0}{\tau}+
\pdrv{{\cal X}^\mu_0}{\sigma}\pdrv{{\cal X}^\nu_0}{\sigma} \right) = 0\ .
\ee
As in the exact non-linear case, if these linearized constraints
are satisfied at the initial moment of time $\tau$ they are
also valid for any $\tau$ for a solution $\chi^{\mu}$ of the dynamical
equations (\ref{2.15}).

The linearized equations can be used to study string motion
in the case where it is far away from the black hole. In this case the
gravitational field can be approximated as follows:
\be\n{2.19}
 ds^2  = -(1-\frac{2M}{R})dT^2+(1+\frac{2M}{R})\, (dX^2+dY^2+dZ^2) 
           -\frac{4J}{R^3}(XdY-YdX)dT\ ,
\ee
where $R^2=X^2+Y^2+Z^2$, and $M$ and $J=aM$ are the mass and angular
momentum of the black hole, respectively. In fact this asymptotic
form of the metric is  valid for any arbitrary stationary localized
distribution of matter, provided that the observer is located far from
it. In agreement with   (\ref{2.19}) the gravitational field
perturbation can be presented as
\be\n{2.20}
q_{\mu\nu}=q_{\mu\nu}^{N}+q_{\mu\nu}^{LT}\, ,
\ee
where the Newtonian and Lense-Thirring parts are
\be\n{2.21}
q_{\mu\nu}^{N}=2\varphi \delta_{\mu\nu}\, ,\hspace{1cm}
\varphi={M\over R}\, , \hspace{1cm}  q_{\mu\nu}^{LT}={4J\over R^3}\,
\delta^0_{(\mu}\, \epsilon_{\nu)\alpha 0 3} X^{\alpha}\, .
\ee
Here $\epsilon_{\alpha\beta\gamma\delta}$ is the antisymmetric symbol.
The Lense-Thirring part $q_{\mu\nu}^{LT}$ of the metric \cite{MTW,ThLe:18} is produced
by the rotation of the source of the gravitational  field and it is
responsible for frame dragging.

\subsection{Newtonian scattering}

In the linear approximation we can study the action of each of the 
parts of the metric perturbations  independently. For the Newtonian
part one has the following expression for the force $f^{\mu}_{N}$
\bean\n{2.22}
f^{0}_{N} &=&2\sinh(\beta)\cosh(\beta)\, ,\varphi_{,1}\\
f^{1}_{N} &=& 0\, , \\
f^{2}_{N} &=& -2\sinh^2\beta\varphi_{,2}\, ,\\
f^{3}_{N} &=& -2\cosh^2\beta\varphi_{,3}\ ,
\eean
and the  constraint equations read
\bean\n{2.26}
\chi^3_{,\tau}-\cosh\beta\chi^0_{,\sigma}+
\sinh\beta\chi^1_{,\sigma} &=& 0 \, ,\\
\chi^3_{,\sigma}-\cosh\beta\chi^0_{,\tau}+
\sinh\beta\chi^1_{,\tau} &=& -2\varphi\cosh^2\beta\ .
\eean

The result $f^1=0$ means that in the first order
approximation the string lies in the $YZ-$plane at any fixed time
$\tau=\mbox{const}$. The dynamical equations (\ref{2.3})
can be easily integrated by using the retarded Green function (see
\cite{DVFr:98}).

 We focus now our attention on the behavior of
$\chi^2$ which describes the string deflection in the plane
orthogonal to its motion. It is easy to show that the asymptotic value
of $\chi^2(\tau\to\infty,\sigma)$ is the same for any fixed
$\sigma$. We denote it $b-\kappa$. The following expression is valid
for $\kappa$
\be\n{2.28}
\kappa=-{1\over 2 \sinh \beta}\, \int_{\Pi_0}\, dX \, dZ f^{2}\, ,
\ee 
where $\Pi_0$ is the  two-dimensional worldsheet swept by the
string in its motion in the background spacetime.
This integral can be easily calculated since 
\be\n{2.29}
\int_{\Pi_0}\, dX \, dZ \varphi_{,Y}=2\pi M\, 
\ee 
is the flux of the Newtonian field strength through $\Pi_0$. Thus we
have
\be\n{2.30}
\kappa=2\pi M \sinh\beta\, .
\ee

\begin{figure}[ht]
\begin{center}
\epsfig{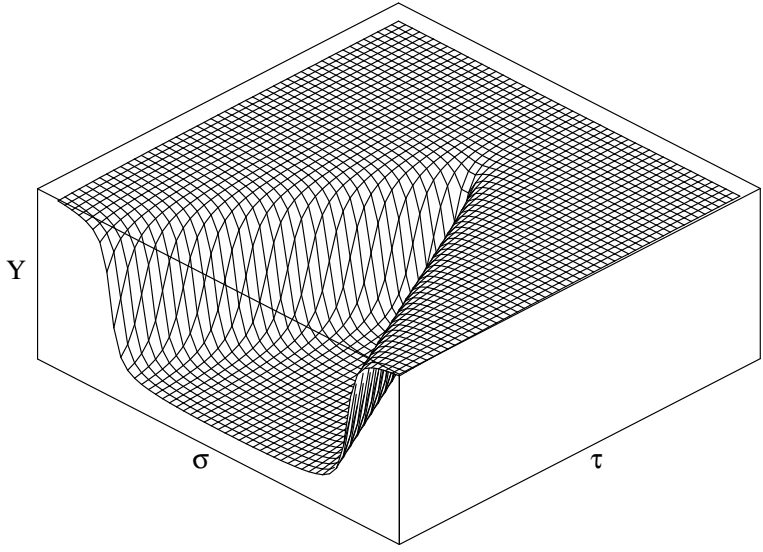}
\caption{$Y-$profile for Newtonian scattering}
\label{cs_Y}
\end{center}
\end{figure}

At any late moment of time $\tau$ the central part of the string is a
straight line moving in the plane $Y=b-\kappa$, while its far distant
parts move in the original plane $Y=b$. These different ``phases'' are
connected by kinks propagating in the direction from the center with
the velocity of light. (For details, see \cite{DVFr:99}). We call
$\kappa$ the {\em displacement parameter}.
Figure~\ref{cs_Y} shows the Y-direction displacement for the Newtonian 
scattering as a function of $\tau$ and $\sigma$.

\subsection{Lense-Thirring scattering }

In the presence of rotation the Lense-Thirring force acting on the
string in the linearized approximation is
\bean\n{2.31}
f^{0}_{LT}  &=& 6J\sinh^2\beta\frac{b (X_0+\tau\sinh\beta)}{{\cal
R}^5} \, ,\\
\n{2.32}
f^{1}_{LT}   &=& 0 \, ,\\
\n{2.33}
f^{2}_{LT}   &=& 2J\sinh\beta\cosh\beta\left(\frac{1}{{\cal R}^3}-
\frac{3\sigma^2}{{\cal R}^5}\right) \, ,\\
\n{2.34}
f^{3}_{LT}   &=& 6J\sinh\beta\cosh\beta \frac{b\sigma}{{\cal R}^5}\, ,
\eean
where ${\cal R}=\sqrt{(X_0+\tau\sinh\beta)^2+b^2+\sigma^2}$.
As before, $f^1=0$.

The constraint equations now take the form
\be\n{2.35}
\chi^3_{,\tau} - \cosh\beta\chi^0_{,\sigma}+\sinh\beta\chi^1_{,\sigma}
= 0\, ,
\ee
\be\n{2.36}
\chi^3_{,\sigma} - \cosh\beta\chi^0_{,\tau}+\sinh\beta\chi^1_{,\tau} 
= -2J\sinh\beta\cosh\beta\frac{b}{{\cal R}^3}\, .
\ee

The dynamical equations can be  solved analytically.
For the initial conditions
\be\n{2.37}
\chi^0|_{\tau=0} = 0\, ,
\ee
\be\n{2.38}
\chi^2|_{\tau=0} =\pdrv{\chi^2}{\tau}|_{\tau=0} = 0\, ,
\ee
\be\n{2.39}
\chi^3|_{\tau=0} =\pdrv{\chi^3}{\tau}|_{\tau=0} =0\, ,
\ee
the solution is
\[
\chi^0 = Jb\sinh\beta\left(\frac{(X_0+\tau\sinh\beta)\sinh\beta
+\sigma}{\Delta_{(+-)}{\cal R}}
+ \frac{(X_0+\tau\sinh\beta)\sinh\beta-\sigma}{\Delta_{(++)}{\cal
R}}\right. \, 
\]
\be\n{2.40}
- \left.\frac{X_0\sinh\beta-(\tau-\sigma)}{\Delta_{(+-)}\rho_-}
- \frac{X_0\sinh\beta-(\tau+\sigma)}{\Delta_{(++)}\rho_+}\right) \, ,
\ee
\be\n{2.41}
\chi^1 = 0\, ,
\ee
\[
\chi^2 = J\sinh\beta\cosh\beta
\left(-\frac{\tau^2\sinh^2\beta+\tau\sigma\sinh^2\beta
+2X_0\tau\sinh\beta+\sigma X_0\sinh\beta+X_0^2  +
b^2}{\Delta_{(++)}{\cal R}}\right. 
\]
\be\n{2.42}
+\frac{-\tau^2\sinh^2\beta+\tau\sigma\sinh^2\beta
-2X_0\tau\sinh\beta+\sigma X_0\sinh\beta-X_0^2 -
b^2}{\Delta_{(--)}{\cal R}}
\ee
\[
+ \left.\frac{X_0\sinh\beta(\tau+\sigma)+X_0^2
+b^2}{\Delta_{(++)}\rho_+} +\frac{X_0\sinh\beta(\tau-\sigma)+X_0^2
+b^2}{\Delta_{(--)}\rho_-}\right)\, ,
\]
\[
\chi^3 = Jb\sinh\beta\cosh\beta\left(\frac{(X_0+\tau\sinh\beta)\sinh\beta
-\sigma}{\Delta_{(++)}{\cal R}}
-\frac{(X_0+\tau\sinh\beta)\sinh\beta+\sigma}{\Delta_{(--)}{\cal R}}\right.
\]
\be\n{2.43}
- \left.\frac{X_0\sinh\beta-(\tau+\sigma)}{\Delta_{(++)}\rho_+} 
+\frac{X_0\sinh\beta-(\tau-\sigma)}{\Delta_{(--)}\rho_-}\right)\, ,
\ee
\be\n{2.44}
\Delta_{\pm\pm} = (X_0\pm(\tau\pm\sigma)\sinh\beta)^2 +
b^2\cosh^2\beta\, ,
\ee
\be
\rho_{\pm} = \sqrt{X_0^2+b^2+(\tau\pm\sigma)^2}\, .
\ee

Using these relations it is possible to see that the displacement
parameter $\kappa=\lim_{\tau\to\infty}\, \chi^2(\tau,\sigma=0)$
vanishes for this solution. The reason  is the following. Using
(\ref{2.33}) it is easy to present $f_{LT}^2$ as
\be\n{2.44a}
f_{LT}^2=2J\sinh\beta\, \cosh\beta {\partial\over \partial
\sigma}\left({\sigma\over {\cal R}^3}\right)\, .
\ee
The asymptotic displacement is given by (\ref{2.28}). The integral
over $Z$ or, equivalently, the integral over $\sigma$ from
$f_{LT}^2$, which is a total derivative over $\sigma$, reduces to
boundary terms $\sigma/{\cal R}^3$ at $\sigma=\pm\infty$, which
vanish.

\begin{figure}[ht]
\begin{center}
\epsfig{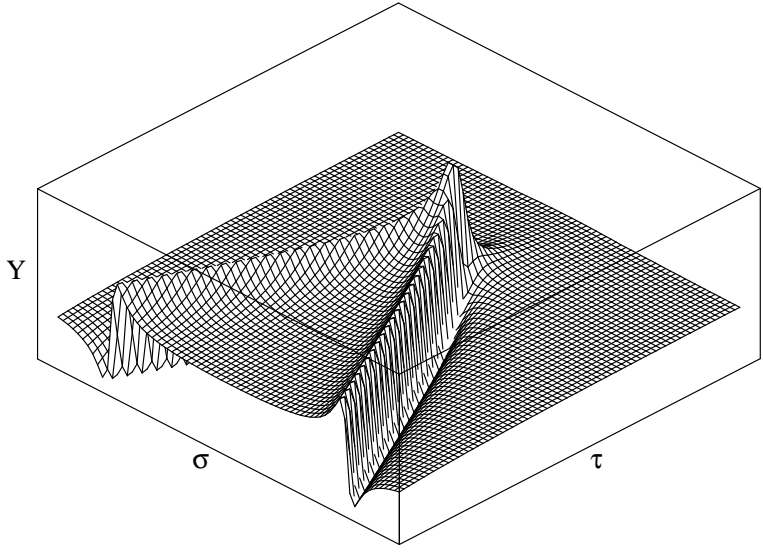}
\caption{$Y-$profile for Lense-Thirring scattering}
\label{cs_lt}
\end{center}
\end{figure}

Figure~\ref{cs_lt} shows the $Y-$direction displacement as a
function of $\tau$ and $\sigma$ for the weak field scattering. It
should be emphasized that the scale of structures for
Lense-Thirring scattering is much smaller than that for
Newtonian scattering. This can be easily seen if we compare the
Newtonian force $f_N\sim M/{\cal R}^2$ with the Lense-Thirring force
$f_{LT}\sim J/{\cal R}^3$
\be
{f_{LT}\over f_N}\sim {J\over M{\cal R}}\le {J\over Mb}\, .
\ee
For the scattering by a rotating black hole $J=aM$, where $|a|/M \le
1$ is the rotation parameter. Hence
\be
{f_{LT}\over f_N} \le {M\over b}\, 
\ee
which is small for the weak field scattering. For this reason, in the
weak field regime the string profiles for the prograde or
retrograde  scattering by rotating black hole do not greatly
differ from the profiles for the scattering by a non-rotating black
hole of the same mass. The situation is 
different for strong-field scattering: the displacement parameter
is different for prograde and retrograde scattering, and for near-critical
scattering the profiles of the kinks contain visible
structure produced by effects connected with the rotation of the
black hole.

\section{String scattering by a Kerr black hole}
\setcounter{equation}0

\subsection{String in the Kerr spacetime}

Our aim is to study a string motion in the Kerr spacetime. The Kerr 
metric in  Boyer-Lindquist coordinates $(t,r,\theta,\phi)$ has 
the form
\bean\n{3.1}
ds^2 &=&  -(1-\frac{2Mr}{\Sigma})dt^2+\frac{\Sigma}{\Delta}dr^2 +
         \Sigma d\theta^2 + \frac{A\sin^2{\theta}}{\Sigma}d\phi^2 -
         \frac{4aMr\sin^2{\theta}}{\Sigma}dtd\phi \ ,\\
\Sigma &=& r^2+a^2\cos^2{\theta} \ ,\\
\Delta &=& r^2 - 2Mr + a^2 \ ,\\
A &=& (r^2+a^2)^2 - a^2\Delta \sin^2{\theta} \ ,
\eean
where $M$ is the mass of the black hole, and $J=aM$ is its angular momentum 
($0\le a\le M$). At far distances this metric has the asymptotic form
(\ref{2.19}).

In order to be able to deal with the case where part of the string
crosses the event horizon for the numerical simulation we adopt the
so-called Kerr  (in-going) coordinates $(\td{v},r,\theta,\td{\phi})$
\[
ds^2 = -(1-\frac{2Mr}{\Sigma})d\td{v}^2 + 2d\td{v}dr - 
         \frac{4aMr\sin^2{\theta}}{\Sigma}d\td{v}d\td{\phi} -
         a\sin^2{\theta}drd\td{\phi}
\]	 
\be \n{3.5}
+ \Sigma d\theta^2 + \frac{A\sin^2{\theta}}{\Sigma}d\td{\phi}^2\ ,
\ee
with
\bean\n{3.6}
\td{v} &=& t+r+M\ln|{\Delta\over 4M^2}| + \frac{M^2}{\sqrt{M^2-a^2}}\ln
      \left|\frac{r-M-\sqrt{M^2-a^2}}{r-M+\sqrt{M^2-a^2}}\right| + M\ ,\\
\td{\phi} &=& \phi+\frac{a}{2\sqrt{M^2-a^2}}\ln
         \left|\frac{r-M-\sqrt{M^2-a^2}}{r-M+\sqrt{M^2-a^2}}\right|\ .
\eean
The metric (\ref{3.1}) has the asymptotic form
\be\n{3.7}
ds^2 = -dt^2 + dr^2 + r^2(d\theta^2 + \sin^2\theta d\phi^2) + 
       \frac{2M}{r}(dt^2+dr^2) - \frac{4aM}{r}\sin^2\theta dtd\phi\ .
\ee
Let us introduce new ``quasi-Cartesian'' coordinates
\bea\n{3.8}
T &=& t\ ,\\
X &=& R\sin{\theta}\cos{\phi}\ ,\\
Y &=& R\sin{\theta}\sin{\phi}\ ,\\
Z &=& R\cos{\theta}\ , 
\eea
where 
\be\n{3.8b}
r=R+M\ .
\ee
One can easily check that the metric (\ref{3.7}) in the quasi-Cartesian coordinates (\ref{3.8})
has the asymptotic form (\ref{2.19})

It should be emphasized that without the shift (\ref{3.8b}) of the radial coordinate one does
not recover (\ref{2.19}). The reason for this is easy to understand if one considers the same
problem for the Schwarzschild geometry. The asymptotic limit of (\ref{2.19}) with $a=0$ can 
be found in the isotropic coordinates with $r = \tilde{R}(1+M/2\tilde{R})^2$ in which the 
spatial part of the metric is conformal to the flat metric. Asymptotically, it is sufficient
to use (\ref{3.8b}) which is the leading part of $r$ at large $\tilde{R}$.

In what follows we shall use these quasi-Cartesian coordinates
for representing the position and the form of the string in the Kerr
spacetime even if we are not working in the weak field regime.
It should be emphasized that since the space in the Kerr
geometry is not flat, plots constructed in these coordinates do not
give a ``real picture''. This is a special case of a general problem
of the visualization of physics in a 4-dimensional curved spacetime.

\subsection{Initial and boundary conditions}
\label{sec3.2}

In an ideal situation, we would study scattering by initially placing
our infinitely long string at spatial infinity,
where the spacetime is flat and the straight string can be described in
simple terms.
In a numerical scheme, however, the string cannot be infinitely long
and we must start the simulation at a finite distance  from the black
hole. We discuss here the initial and boundary conditions used for the
simulations.

\begin{figure}[ht]
\begin{center}
\epsfig{file=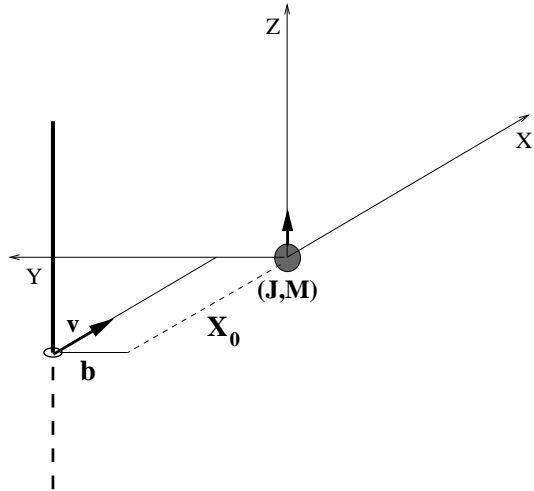, width=7cm}
\caption{Initial setup of the scattering experiment}
\label{InitSetup}
\end{center}
\end{figure}

In studying the scattering of a straight string, we
consider the special case where the string is initially parallel to the
axis of rotation of the black hole. We use Cartesian coordinates
in the asymptotic region so that $X$-axis  coincides with the
direction of motion and the $Z$-axis is along the string
(see Fig.~\ref{InitSetup}).

We consider a string segment of length $L\sim 10^4M$ at a
time $\tau_0$ and an
initial distance $X_0$ from the black hole.
In order to keep accuracy high and yet prevent the calculation
time from being inordinately large, we do not evolve the straight string
numerically from this initial position. Instead, we
use the weak-field approximation to describe the string
configuration at a later time
$\tau_s$ where the distance $X_s$ is closer to the black hole, $|X_s|\ll |X_0|$
~\footnote{ In our simulations we
put $X_0 = -10^6 M$ and $X_s = -500 M$  }.

\begin{figure}[ht]
\begin{center}
\epsfig{file=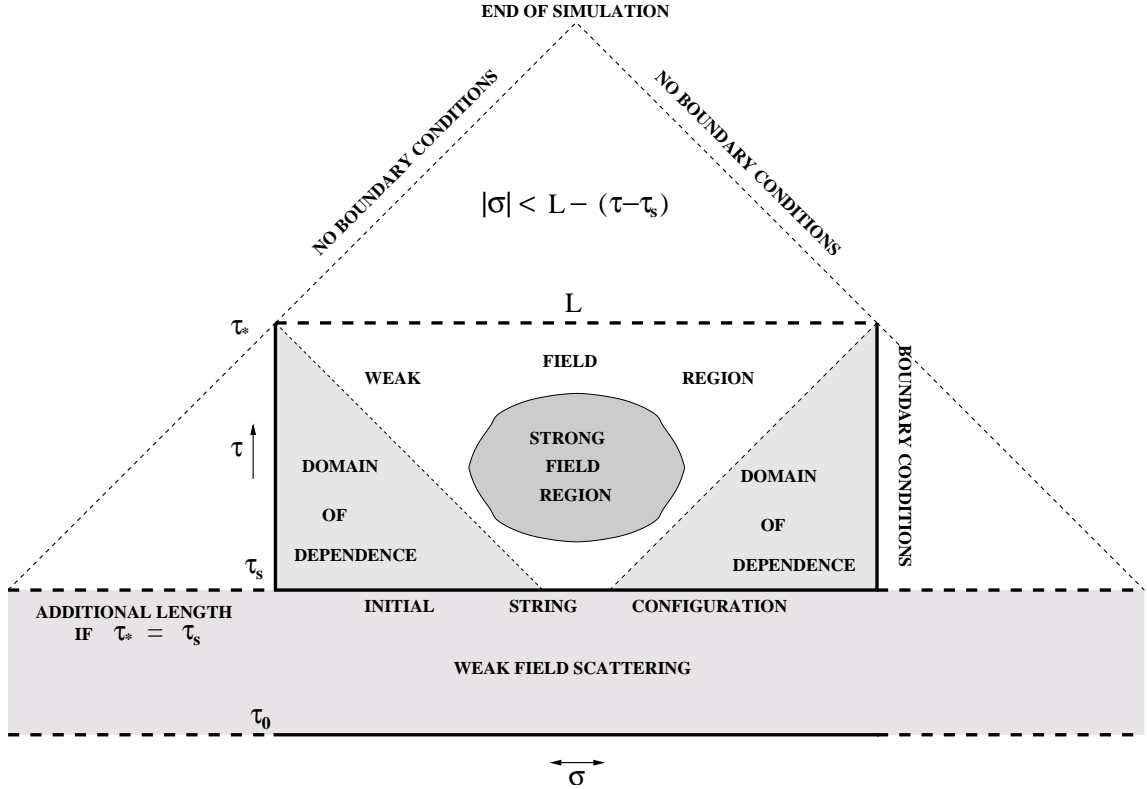, width=\textwidth}
\caption{Scheme of  time domains for scattering problem}
\label{f02}
\end{center}
\end{figure}

Given a sufficiently long string segment, the boundary points move
at a great distance
from the black hole, so that their evolution can be described by the
weak field approximation until information about the
interaction with the black hole reaches the boundary. We denote this
time as $\tau_*$. Starting from this moment we solve the dynamical
equations in the region $|\sigma|< L-(\tau-\tau_*)$ (see
figure~\ref{f02}). The larger is the initial value $L$ the longer one can
go in $\tau$ in the simulation. We choose $L$ to be large enough to
provide the required accuracy in the determining the final scattering
data.

Since the boundary conditions found from the weak field regime calculations
are not exact there will be disturbance created by this effect.
In some cases, when it is important to exclude
these disturbances, we used a modified scheme of calculations in which 
no boundary conditions are used from the very beginning of the simulation.
The price for this is that one needs to take longer initial size of 
$L$ which in turn increases the computational time.

\subsection{Solving dynamical equations and constraints}

Using the initial conditions at $\tau_s$ and boundary conditions we
solve numerically the dynamical equations (\ref{2.8}).
Since the equations are
of second order, we use the weak-field approximation to get initial data at
$\tau_s$ and $\tau_s+\Delta\tau$ in order to completely specify the initial
value problem.  The numerical scheme uses second-order finite differences and
evolves the string configuration using an implicit scheme.  Because of the
symmetry $\sigma\to -\sigma$ it is sufficient to evolve only half of the string
worldsheet, $\sigma\le 0$, and use a reflecting
boundary condition at the string midpoint. 
The numerical grid is non-uniform with the denser part following the motion
of the kink.
We used the
constraint equations (\ref{2.9}-\ref{2.10}) for an independent check of the
accuracy of the calculations.  More details on the numerical scheme
can be found in the Appendix.

\section{String profiles for strong field scattering}
\setcounter{equation}0

\subsection{General picture}

\begin{figure}
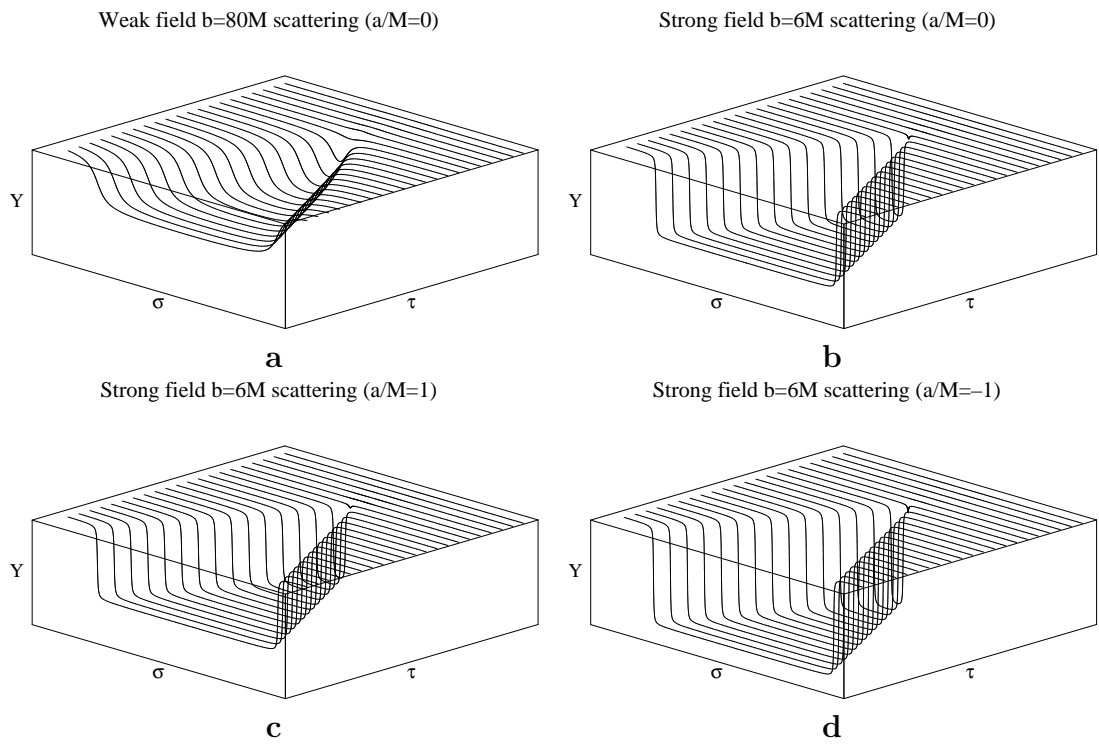

\begin{tabular}{cc}
\epsfig{file=scat_i80p0_a0p0_b1p0_tsk.eps, width=7cm}
&
\epsfig{file=scat_i6p0_a0p0_b1p0_tsk.eps, width=7cm}\\
{\bf a} & {\bf b} \\
\epsfig{file=scat_i6p0_a1p0_b1p0_tsk.eps, width=7cm}
&
\epsfig{file=scat_i6p0_a-1p0_b1p0_tsk.eps, width=7cm}\\
{\bf c} & {\bf d} \\
\end{tabular}
\caption{Displacement in $Y-$ direction as a function of
$(\tau,\sigma)$ for given velocity $v/c=0.762$ in the weak (a) and
strong (b--d) field regime.}
\label{3d}
\end{figure}

Figure~\ref{3d} demonstrates general features of straight string scattering,
here for a string with velocity $v=0.762$ ($\beta=1$).  The four panels show the
displacement in the $Y-$direction as a function of internal coordinates
$(\tau,\sigma)$ for weak and strong field scattering.

Figure~\ref{3d} (a) shows the scattering for an impact parameter $b=80\, M$.
The value of the displacement parameter at the largest $\tau$ shown is
$\kappa=7.23M$.  This value as well as the form of the kinks are in a very good
agreement with the weak-field scattering results.
Plots for prograde and retrograde scattering at $b=80\, M$ are
virtually indistinguishable from Fig.~\ref{3d} (a).

Figures~\ref{3d} (b--d) show the scattering for  $b=6\, M$.
Qualitatively these plots are similar to weak-field scattering. The main
differences are the following: (1) The value of the displacement parameter
$\kappa=11.03M$ for the non-rotating
black hole is significantly larger then $\kappa=7.23$ for $b=80\, M$.
(2) Effects of rotation are
more pronounced. The displacement $\kappa$ for retrograde scattering is
essentially larger than $\kappa$ for the prograde scattering. (3) The
string profile after scattering is more sharp, the width of the kinks
is visibly smaller that for weak-field scattering.

\subsection{``Real-time profiles'' of the string for strong-field scattering}

\begin{figure}
\begin{tabular}{c}
\epsfig{file=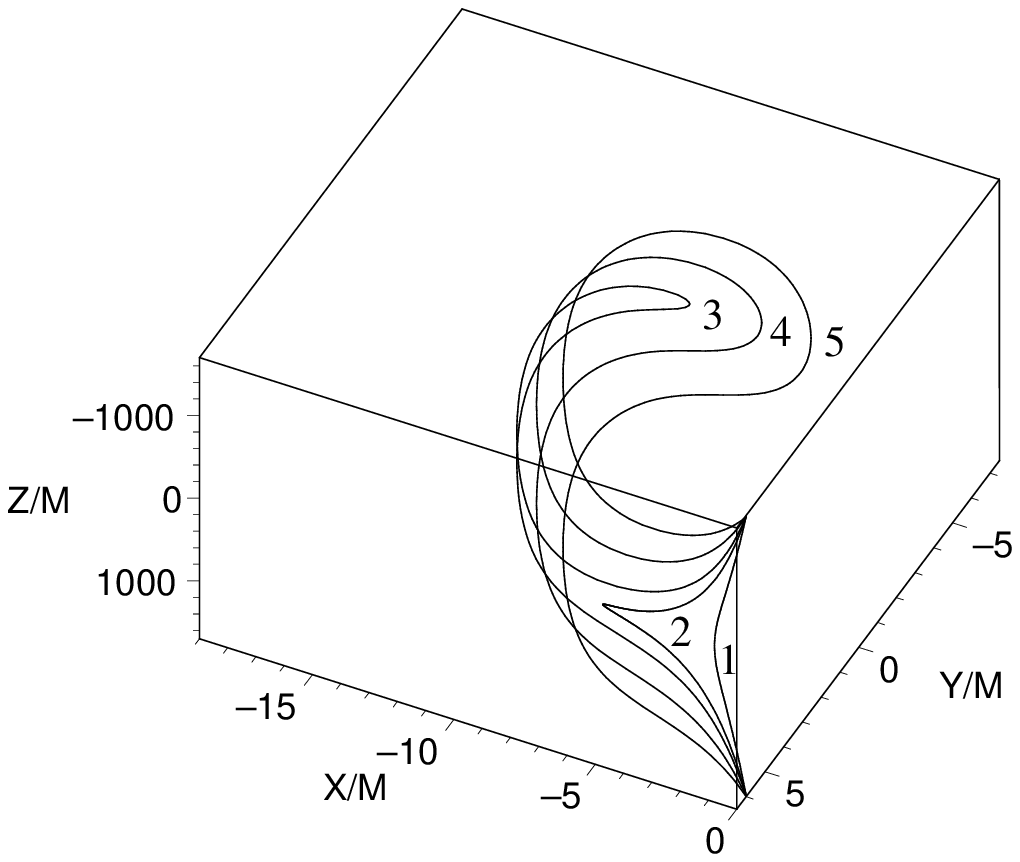, width=8cm}
\end{tabular}
\hfill
\begin{tabular}{c}
\epsfig{file=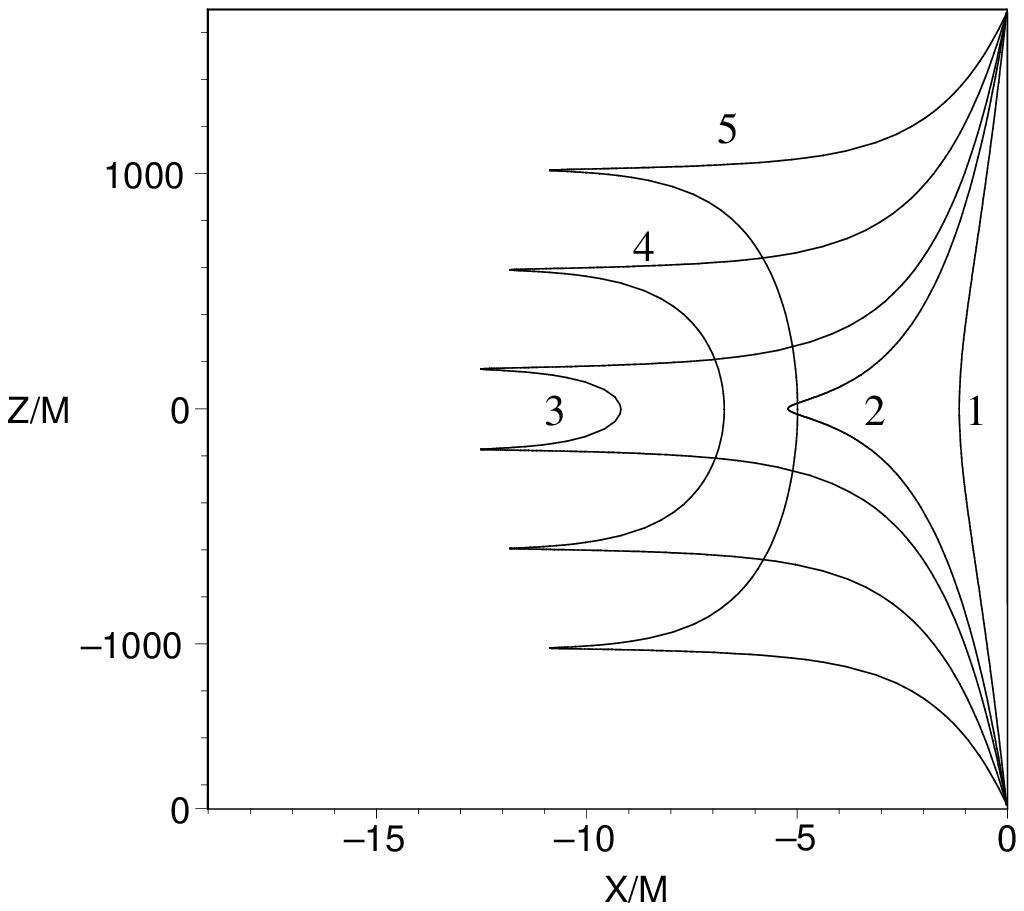, width=4.5cm}\\
\epsfig{file=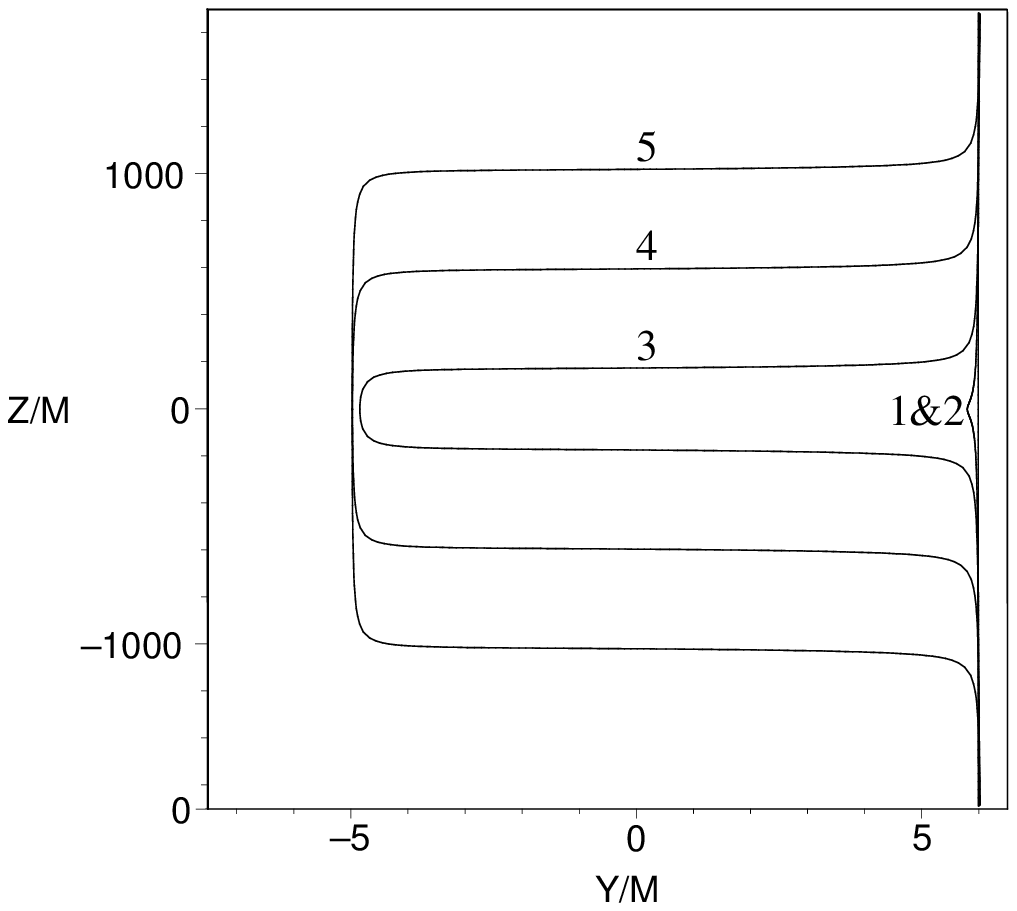, width=4.5cm}
\end{tabular}
\caption{``Real-time profiles'' of the string and their $XZ-$ and
$YZ-$ projection for the strong field scattering by the Schwarzschild
 black hole.}
\label{RT_S}
\end{figure}

\begin{figure}
\begin{tabular}{c}
\epsfig{file=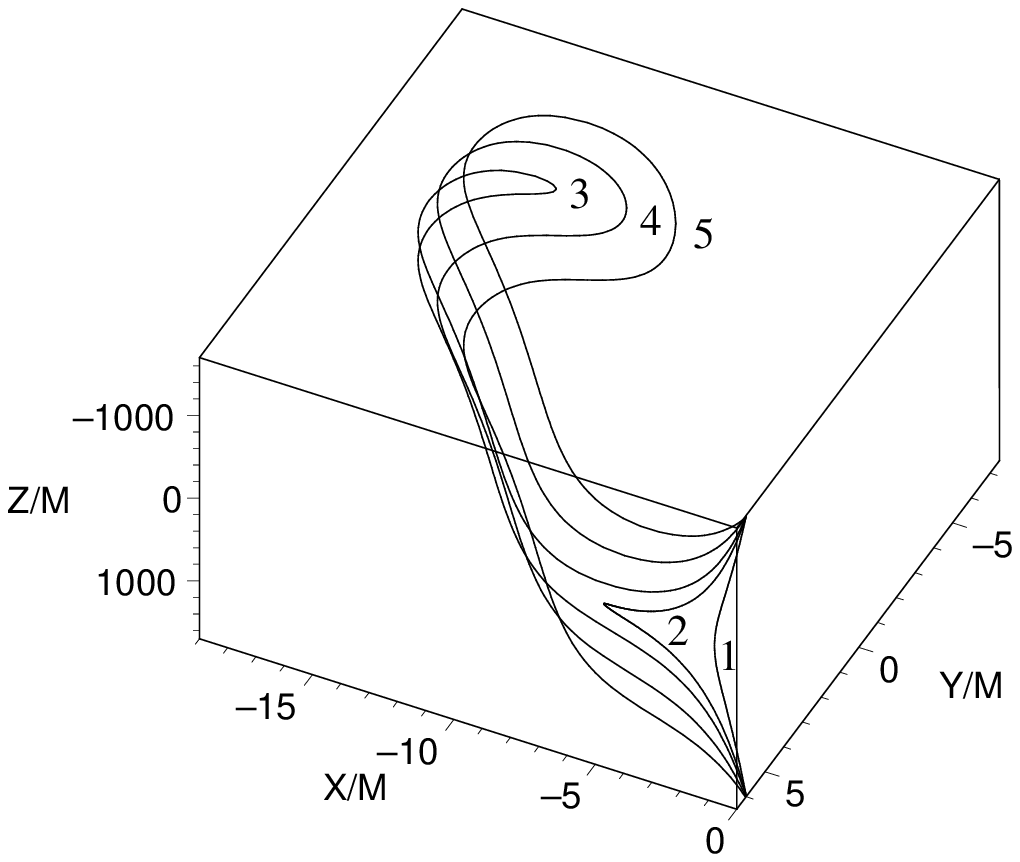, width=8cm}
\end{tabular}
\hfill
\begin{tabular}{c}
\epsfig{file=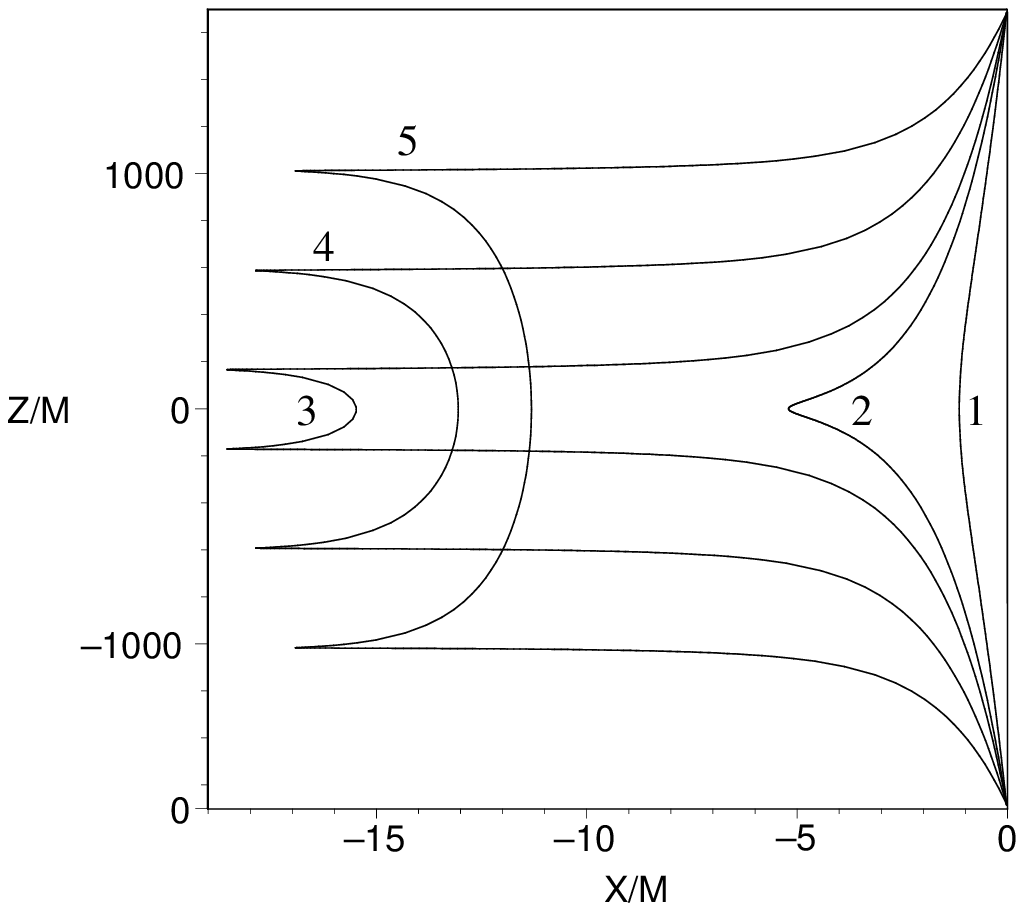, width=4.5cm}\\
\epsfig{file=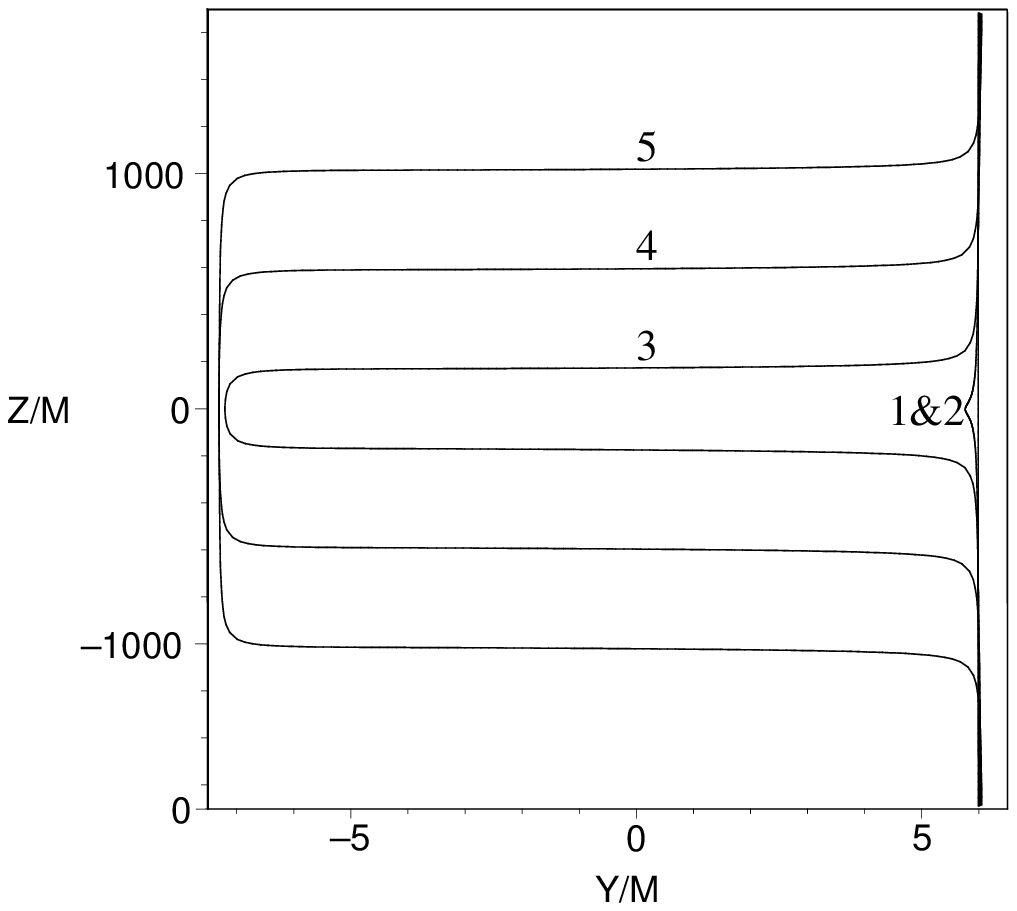, width=4.5cm}
\end{tabular}
\caption{``Real-time profiles'' of the string and their $XZ-$ and
$YZ-$ projection for the strong field retrograde scattering by the 
extremal Kerr black hole.}
\label{RT_K-}
\end{figure}

\begin{figure}
\begin{tabular}{c}
\epsfig{file=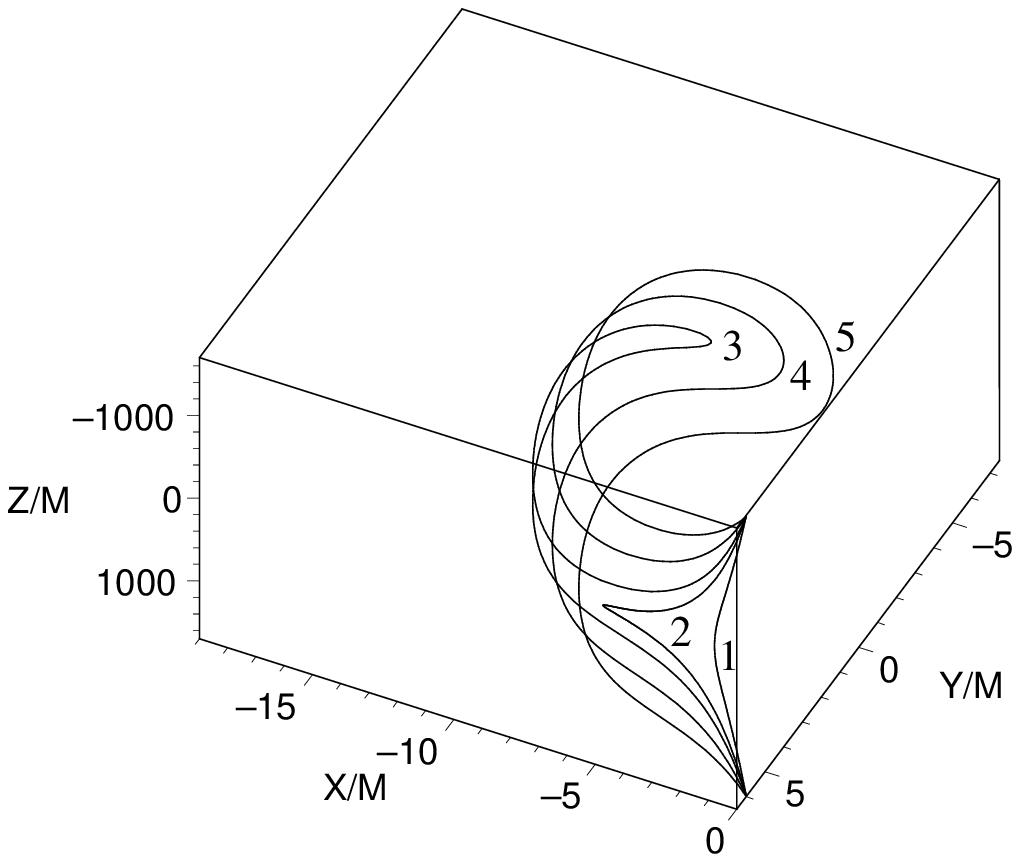 , width=8cm}
\end{tabular}
\hfill
\begin{tabular}{c}
\epsfig{file=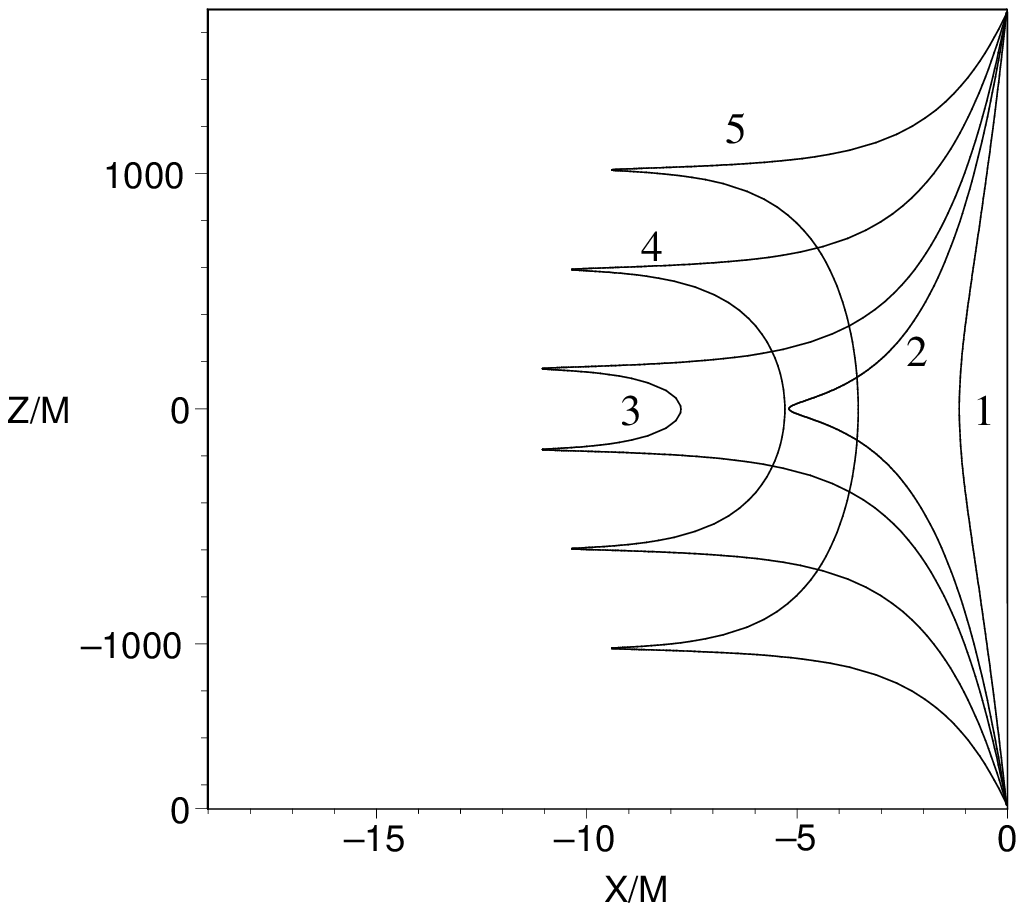, width=4.5cm}\\
\epsfig{file=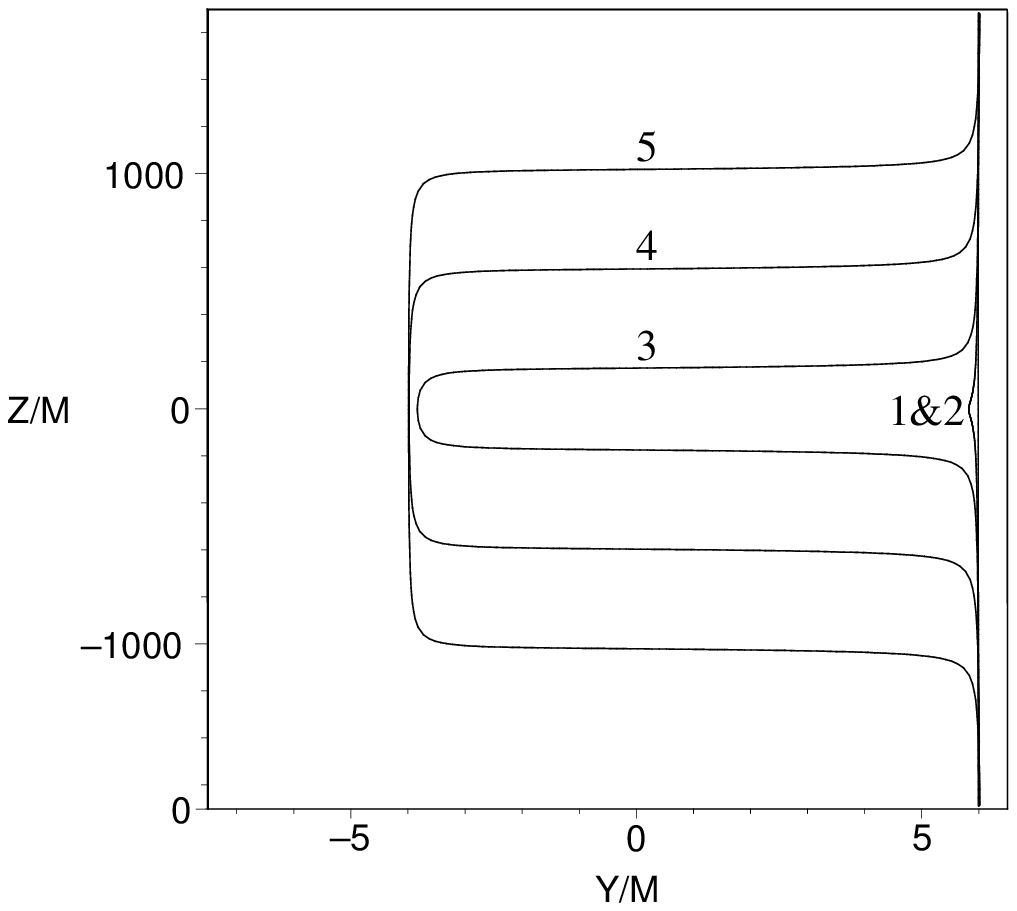, width=4.5cm}
\end{tabular}
\caption{``Real-time profiles'' of the string and their $XZ-$ and
$YZ-$ projection for the strong field prograde scattering by the 
extremal Kerr black hole.}
\label{RT_K+}
\end{figure}

It should be emphasized that figure~\ref{3d} give an accurate impression of the
displacement effect, but they give a distorted view of the real form of the
string.  The reason is evident.  The grid imposed on the worldsheet is
determined by the choice of $(\tau,\sigma)$ coordinates.  But a $\tau={\rm const}$
section differs from a time $T={\rm const}$ section in the laboratory slice of
spacetime.  The position of the string at given time $T$ can be found by using
the functions $T(\tau,\sigma)$, $X(\tau,\sigma)$, $Y(\tau,\sigma)$,
$Z(\tau,\sigma)$, to find functions $X(T,Z)$ and $Y(T,Z)$.  For fixed $T$ these
functions determine a position of the string line in 3-space.

Figures~\ref{RT_S}, \ref{RT_K-} and \ref{RT_K+} show a sequence of
``real-time'' profiles for strong-field string scattering at five
different coordinate times $T$. We ordered them so that the larger
number labeling the string corresponds to later time.
In figure~\ref{RT_S} the black hole is non-rotating.
Figures~\ref{RT_K-} and \ref{RT_K+} show the ``real-time'' profiles
for retrograde and prograde scattering for a maximally rotating black
hole, respectively. Again we can see that due to frame dragging, the
distortion of the string is more pronounced for retrograde scattering.

\section{Late time scattering data}
\setcounter{equation}0

\subsection{Displacement parameter}

\begin{figure}
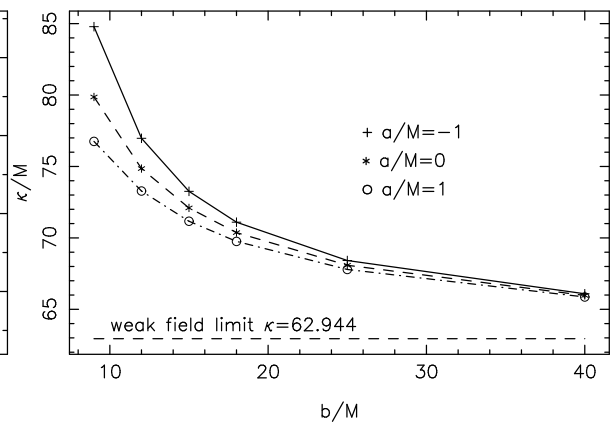

\begin{tabular}{c}
\hfill 
\epsfig{file=k_vs_i_v1p0f.eps, width=8cm}\\
\epsfig{file=k_vs_i_v2p0f.eps, width=8cm}\hfill
\epsfig{file=k_vs_i_v3p0f.eps, width=8cm}
\end{tabular}
\caption{Displacement parameter $\kappa$ as a function of the impact
parameter $b$ for different velocities $v$.}
\label{disp}
\end{figure}

Figure~\ref{disp} shows the dependence of the displacement parameter $\kappa$ on
the impact parameter $b$ for different velocities $v$.  For  a given value of $b$
the curve for $a/M=-1$ always lies higher than the curves for more positive 
values of
$a/M$.  This is true for all values of velocity $v$, but this difference is more
pronounced at large velocities.  

The relative location of the curves for given $v$ and $b$ and 
different $a$ can be explained by frame dragging.
Namely,  a string in the retrograde motion is effectively slowed down
when it passes near the black hole and hence it spends more time near
it. As the result its displacement is greater than the displacement
for the prograde motion with the same impact parameter. Let us make
order of magnitude estimation of this effect. For this
purpose we use metric (\ref{2.19}). We focus our attention on the
Lense-Thirring term and neglect for a moment the Newtonian part.
Consider first a point particle moving in this metric with the velocity $v$
and impact parameter $b$. In the absence of rotation it moves with
constant velocity so that $(X=vT, Y=b, Z=0)$ and the proper time $\tau$ is
\be
\tau_0 =\sqrt{1-v^2}\, T\, .
\ee
In the presence of rotation one  has
\be
d\tau^2=(1-v^2)\, dT^2 -{4Jbv\over R^3}\, dT^2\, ,
\ee
where
\be
R^2=b^2+v^2\, T^2\, .
\ee
For large $b$ one can write
\be
\tau =\tau_0 +\Delta\tau\, ,
\ee
where
\be
\Delta\tau = -{2Jbv\over \sqrt{1-v^2}}\, \int_{-\infty}^{\infty}\,
{dT\over (b^2+v^2T^2)^{3/2}}\, =-{4J\over b\sqrt{1-v^2}}\, .
\ee
This quantity $\Delta\tau$ characterize additional time delay for the
motion in the Lense-Thirring field. The characteristic time of motion
in the vicinity of a black hole is $\tau_{int}\sim b/v$. Thus we have
\be
{\Delta\tau\over \tau_{int}}\sim -{4Jv\over b^2\sqrt{1-v^2}}\, .
\ee
One can expect that a similar delay takes place for a string motion.
As a result of being longer close to the black hole the string
has a larger displacement by the value $\Delta \kappa$ such that
\be
{\Delta\kappa\over \kappa}\sim {\Delta\tau\over \tau_{int}}\sim
-{4Ma\over b^2}\sinh \beta \, . 
\n{5.7}
\ee
Qualitatively this relation explains dependence of $\kappa$ on $a$ and
$\beta$ presented in figure~\ref{disp}.

More generally, if we look at the $\alpha=a/M$ in the metric as a parameter, we
can use perturbation theory to calculate the effect of rotation.
The solution ${\cal X}^\mu$ can be expanded in powers of $\alpha$
\be
{\cal X}^\mu = \up{0}{\cal X}{\mu}+\alpha\up{1}{\cal X}{\mu}
                +\alpha^2\up{2}{\cal X}{\mu} + \dots\ ,
\ee
where $\up{0}{\cal X}{\mu}(\tau,\sigma)$ is the solution for the 
non-rotating case and 
$\up{i}{\cal X}{\mu}(\tau,\sigma)$ are the perturbation corrections.

Numerical calculations show that the first two terms in the perturbation series 
give very good approximation to the solution as long as we are  not
close to critical scattering. As a consequence 
\be\n{5.9} 
\kappa = \kappa_0 - \kappa_1\alpha - \kappa_2\alpha^2\ ,
\ee
where $\kappa_0$ is the $Y$ displacement for the non-rotating black hole, and 
$\kappa_1$,$\kappa_2$ are defined simply as
\be
\kappa_{1,2} = \lim_{\tau\to\infty} \up{1,2}{\cal X}{2}(\tau,0)\ .
\ee
In the above $\up{1,2}{\cal X}{2}$ denotes the $Y$ coordinate corrections.
The equations of motion for $\up{1,2}{\cal X}{\mu}$ are linear and 
$\up{1,2}{\cal X}{\mu}$ can be obtained in one run together with 
$\up{0}{\cal X}{\mu}$.

In order to determine the coefficients in equation~(\ref{5.9}) we need 
to extrapolate the relevant data obtained during the simulation to infinity.
One possibility is to numerically advance the solution very far from the black hole and simply
take the value at the end of the simulation as our estimate. The problem with this approach is that
to advance very far by maintaining high accuracy of the solution
is a very lengthy process, feasible only for larger impact parameters (since the grid can be
sparser).

Our approach was to advance to a moderate distance ($X_{\rm max}\approx 8000M$) keeping high accuracy of the
solution (judged by the constraint equations).
Then we fit the calculated data by the function
\be
\kappa(0,\tau) = \kappa + k(\tau-\tau_{sh})^\gamma\ ,
\ee
where $\kappa$, $k$, $\tau_{sh}$, $\gamma$ are the parameters to be fitted.
We take our first data point at $X=1000M$.
To see how consistent this procedure is and to estimate the errors
we perform three fits with different set of data points.
The first set uses all the data points,
the second one uses only the first half,
and the third one uses only the last half of the data points.
For this particular problem the disturbances from the boundary have adverse effect
to the fit therefore we use the null boundary approach.

Our experience shows that the fitting procedure for $\kappa$ works much better 
for larger velocities and larger impact parameters.
For the velocities $v/c = 0.762$, $0.964$, $0.995$ the estimated values of 
$\kappa$ from all three fits are consistent and differ by no more than 3\%
\footnote{For $b\ge12M$ the results from the three fits differ by less than 1\%.}.
This fact gives us reasonable confidence that the plotted values of $\kappa$ 
indeed represent the true values.

Unfortunately the situation is not so good for slower velocities. For example for 
the case $v/c=0.462$ the slope of the 
fitted function $\kappa(0,\tau)$ is almost constant during the later phases
of the simulation and fits using different sets of data points yield 
inconsistent (and improbable) values of $\kappa$.
We expect the behavior of $\kappa(0,\tau)$ to change but we did not see it in our data 
even after driving the simulation four times further then for the other velocities 
($X_{\rm \max}\approx 32,000M$).
Again, this behavior is more pronounced for smaller impact parameters $b$.

\begin{figure}[!htb]
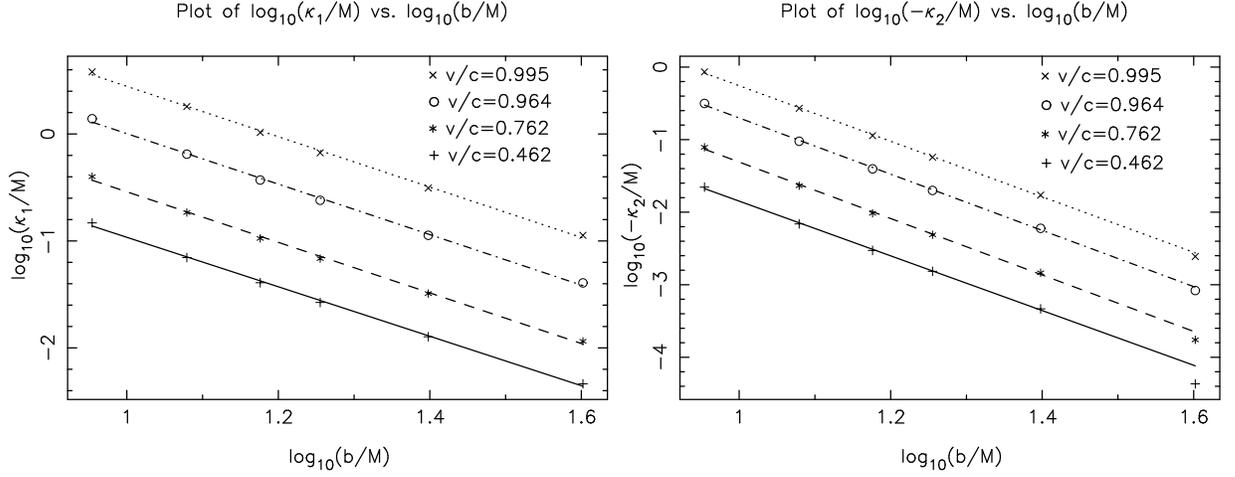

\begin{tabular}{c}
\epsfig{file=k1_vs_i.eps, width=8cm}
\hfill
\epsfig{file=k2_vs_i.eps, width=8cm}\\
\end{tabular}
\caption{Plots of $\kappa_1$ and $\kappa_2$ for different velocities $v$
and impact parameters $b$. The straight lines represent a linear fit.}
\label{kappa_12}
\end{figure}
\begin{figure}[!htb]
\begin{center}
\epsfig{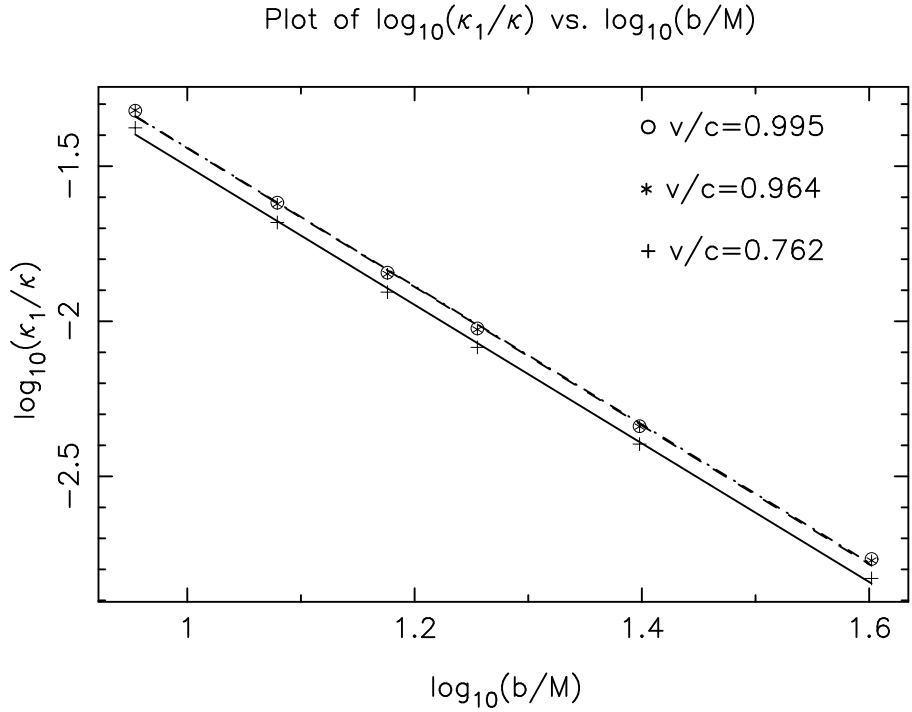}
\caption{Plot of $\kappa_1/\kappa$ for different velocities $v$ 
and impact parameters $b$. The straight lines represent a linear fit.}
\label{k1_over_k}
\end{center}
\end{figure}

To obtain the values for $\kappa_1$ and $\kappa_2$ we use the same method as for
$\kappa$.
We did not encounter any difficulties in estimating the values of $\kappa_1$ and
$\kappa_2$ for any of the four velocities.
The trend here is the opposite of the one for $\kappa$ -- the fits work
best for smaller velocities and smaller impact parameters.
The fits with different set of data points are typically less than $1$\% appart.

Figure~\ref{kappa_12} shows the results in a log-log plot.
Figure~\ref{k1_over_k} shows the plot of $\kappa_1/\kappa$.
The straight lines shown are obtained from least square fit.
Note that the plots for the velocities $v/c=0.964$ and $v/c=0.995$
lie on top of each other.

In the interval shown $\kappa_1$ and $\kappa_2$ 
can be reasonably approximated by a functions of the form
\be
\n{5.12}
\kappa_{1,2} \approx A_{1,2}(v)\left(\frac{1}{b}\right)^{\lambda_{1,2}}\ .
\ee
Similarly for $\kappa_1/\kappa$,
\be
\n{5.12b}
\frac{\kappa_1}{\kappa} \approx A_3(v)\left(\frac{1}{b}\right)^{\lambda_3}\ .
\ee
The values for $A_{1,2,3}$ and $\lambda_{1,2,3}$ obtained from a linear fit are shown
in table~\ref{table1}.
By comparing~(\ref{5.12b}) with (\ref{5.7}) one can conclude that the numerical value
$\lambda_3=2.23$ is close to the value $2$, which enters the relation (\ref{5.7}).

Note that the last data point on the plot for $\kappa_2$ is a bit out of a straight line. 
We are not sure if this is a genuine feature or simply inaccurate data points (e.g., due to
round-off errors)\footnote{The discrepancies are larger than those which could result 
from the extrapolation procedure.}.
We did not include these data points into calculation of the values in table~\ref{table1}.

\begin{table}[!htb]
\bea
\begin{array}{|c||c|c|c|c|}\hline
 v/c        & 0.462 & 0.762 & 0.964 & 0.995  \\ \hline\hline
A_1         & 22.3  & 66.6  & 231.0 & 631.6  \\ \hline
\lambda_1   & 2.31  & 2.36  & 2.36  & 2.35   \\ \hline
A_2         & -83.9 &-381.4 &-1487.9&-3717.0 \\ \hline 
\lambda_2   & 3.77  & 3.89  & 3.87  & 3.83   \\ \hline
A_3         &  -    &5.42   &6.21   & 6.17 \\ \hline 
\lambda_3   &  -    &2.23   &2.23   &2.23 \\ \hline
\end{array}
\eea
\caption{Values of fitted parameters from equation~(\ref{5.12}) and (\ref{5.12b}) }
\label{table1}
\end{table}

\subsection{Form of the kinks}

\begin{figure}[!htb]
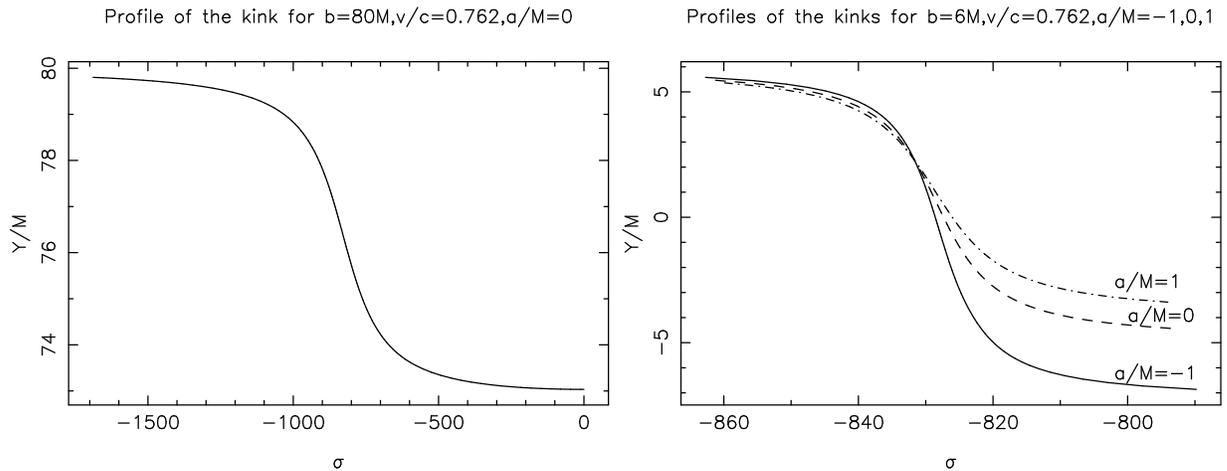

\begin{tabular}{c}
\epsfig{file=kink_b80_i1p0.eps, width=8cm}
\hfill
\epsfig{file=kink_b6_i1p0.eps, width=8cm}\\
\end{tabular}
\caption{Profiles of the kinks for different impact parameters}
\label{kinks}
\end{figure}

Figure~\ref{kinks} shows details of the transition from the ``old phase'' 
to the ``new phase'' for two different impact parameters. 
We see that the width of the kinks differs significantly.
In general the width also depends on the velocity $v/c$..
The dependence of the kink width on velocity and the impact parameter 
can be estimated from the weak field approximation as
\be
w \sim \frac{bc}{v}\ .
\label{n5.2}
\ee

\begin{figure}[!ht]
\begin{center}
\epsfig{file=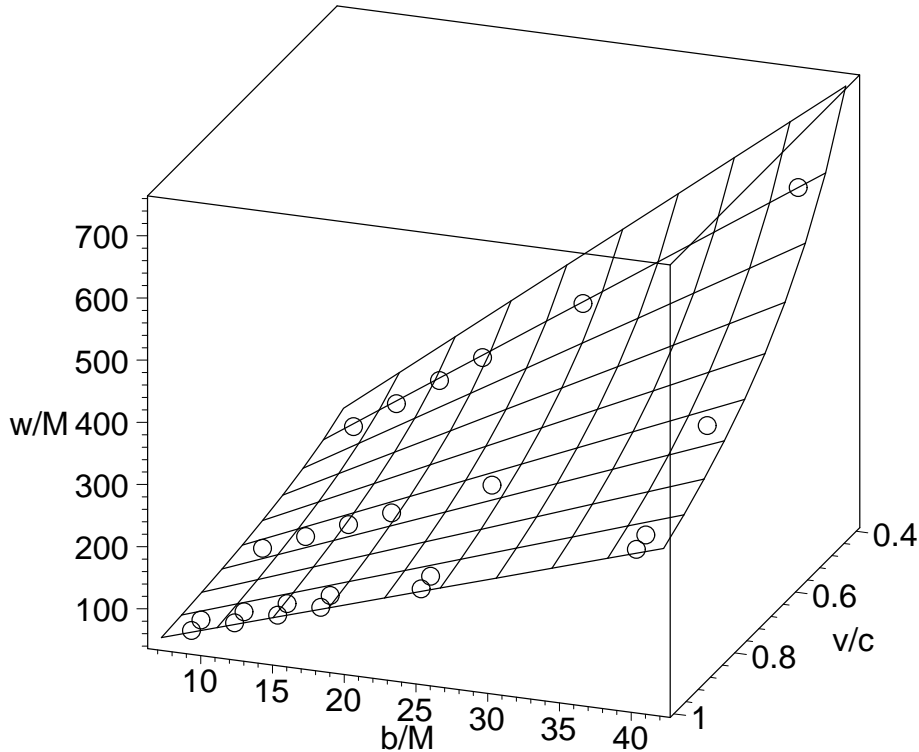, width=12cm}
\caption{Comparison of kink widths for $v/c=0.462,0.762,0.964,0.995$ and 
impact parameters $b/M=9,12,15,18,25,40$ obtained from simulation
(circles) and the analytic approximation~(\ref{5.13}). The data shown are taken
for $a/M=0$.}
\label{kink3d}
\end{center}
\end{figure}

We operationally define the width of the kink to be the width of the peak
 of $dY/d\sigma$ at $1/20$th of its hight.
Figure~\ref{kink3d} shows comparison of the kink widths obtained from 
simulation with a function 

\be
\n{5.13}
w_{\rm analyt} = 7.14\frac{bc}{v}\ .
\ee
The numerical constant was obtained by a fit. The match with the data is very good.

We also noticed that the dependency of the width of the kink 
on the rotation parameter $a$ is rather small.
In general
\be
w_{a/M=1}<w_{a/M=0}<w_{a/M=-1}\ .
\ee
For the parameters shown on figure~\ref{kink3d} the rotation made a difference
 from  $3M$ to $7M$ for the extremal black holes.

\section{Discussion}
\setcounter{equation}0

In this paper we studied the scattering of a long test cosmic string
by a rotating black hole. We demonstrated that qualitatively many
general features of the weak field scattering are present in
scattering of the cosmic string in the strong field regime.
Displacement of the string in the $Y-$direction always has the form
of a transition of the string from the initial phase (initial plane) to the
final phase (final plane which is parallel to the initial one and
displaced in the direction to the black hole by distance
$\kappa$). The boundary between these phases is a kink moving with the
velocity of light away from the center. An important
difference between weak- and strong-field scattering is in the
dependence of $\kappa$ on the impact parameter $b$ and velocity $v$.
In general $\kappa$ for strong field scattering is much
greater that its value calculated by weak-field approximation.
It is also always greater for retrograde scattering than
for a prograde scattering. The explanation of this is quite simple: The
retrograde string spends more time in the vicinity of the black hole
than a prograde one. This effect is a result of the dragging of the
string into rotation by the black hole.

The width of the kink for the
strong field scattering is smaller than for the scattering in the weak 
field regime.

As was demonstrated in previous studies for string scattering by
non-rotating black holes, for any given velocity there exists a critical
value $b_{crit}$ of
the initial impact parameter so that for $b<b_{crit}$ the string is
trapped by the black hole, while for $b>b_{crit}$ it is scattered by
the black hole (and all points on the string reach infinity). In this paper we
focus our attention on scattering in the regime where $b$ is
significantly greater than $b_{crit}$. The study of string capture and
near-critical string scattering requires modification of the
computational program and more time consuming calculations. We are
going to present these results in a subsequent publication.

\vspace{12pt} 
{\bf Acknowledgments}:\ \ The authors are
grateful to Don Page for various stimulating discussions.
This work was partly supported by the Natural
Sciences and Engineering Research Council of Canada.  One of the
authors (V.F.) is grateful to the Killam Trust for its  financial
support. M.S. thanks FS Chia PhD Scholarship.
Our work made use of the infrastructure and resources of MACI (Multimedia Advanced
Computational Infrastructure), funded in part by the CFI (Canada Foundation for Innovation),
ISRIP (Alberta Innovation and Science Research investment Program), and the Universities of
Alberta and Calgary.

\appendix

\section{Numerical scheme details}
\setcounter{equation}0

In section \ref{sec3.2} we briefly described the numerical scheme we adopted.
In this appendix we present some more details.
Figure \ref{f02} shows the overall structure of the time domain.

The numerical grid in the sigma direction is non-uniform. 
This is very important since there are regions where the string is relatively straight and regions
with high curvature. By far the numerically most ``sensitive'' region is the kink.
Therefore we make the grid denser in the vicinity of the kink. The length of the dense zone
is chosen to be approximately twice the kink width\footnote{The grid within each zone is uniform}.
The steeper the kink the denser the grid must be. Typically, the dense zone is $3$--$30$ times
denser then the rest of the grid.

The time step $\Delta\tau$ must not exceed the smallest distance between any two grid 
points $\Delta\sigma$. A convenient choice turns out to be 
\be
\Delta\tau = \Delta\sigma\sqrt{1-v^2/c^2}\ .
\ee
This means that dense grid also implies small time step which in turn makes the 
simulation time longer.
The formula~(\ref{n5.2}) explains why it is more difficult to deal with small impact
parameters and relativistic velocities.

\begin{figure}[!h]
\begin{center}
\epsfig{file=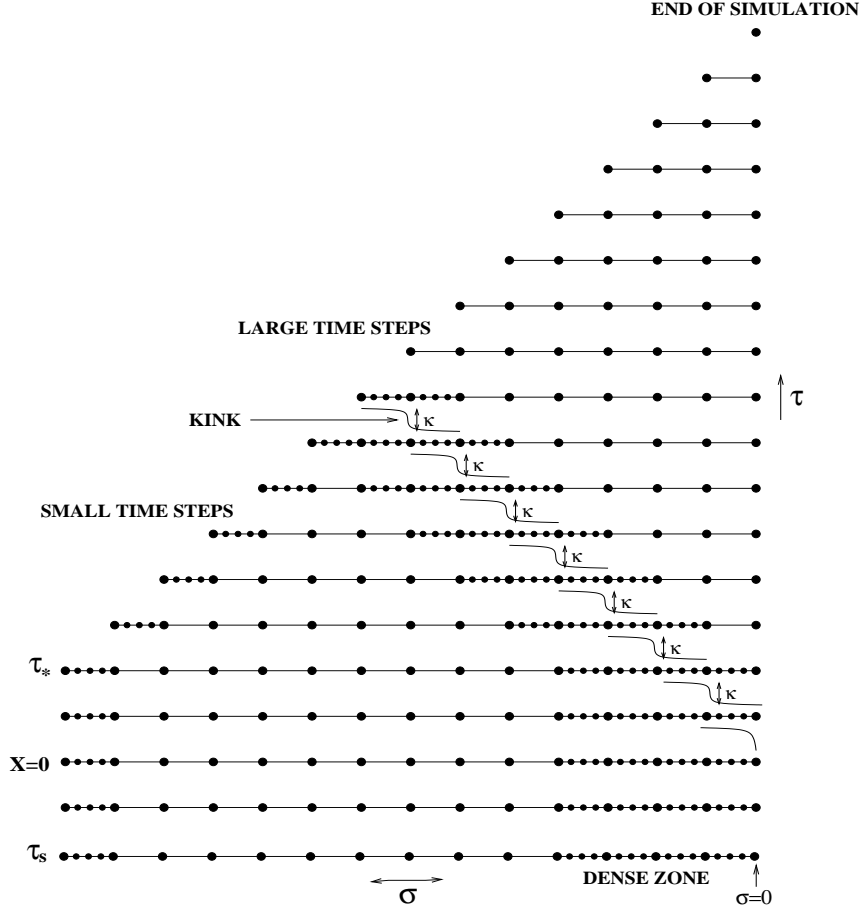, width=0.75\textwidth,height=12cm}
\caption{Structure of the numerical grid}
\label{grid}
\end{center}
\end{figure}

Since the kink is moving away from the central point $\sigma=0$ the dense zone must move 
with it. Therefore we regularly check the position of the kink and adjust the grid 
when the kink reaches the boundary of the dense zone. This procedure is illustrated in figure
\ref{grid}. Note that we introduced similar (but much shorter) dense zone at the edge
of the string. This is to prevent from ``cutting'' the string too quickly (every time 
step we lose one grid point). 
When we change the the grid structure we must use interpolation to obtain the variable values 
at the new grid points. In our case we used cubic spline which proved to work very well.

At certain stage the two dense zones merge and eventually they disappear altogether (which 
means that the kink passed the edge of our numerical domain). At this point we can increase 
$\Delta\tau$ to reflect the sparser grid.

Since our problem is naturally formulated on rectangular grid the finite difference method is
used to discretize the equations of motion.
To approximate the first and second derivatives we use standard second order centered formulas,
e.g.,
\be
(\pa_{\tau}{\cal X}^\mu)_{i,j} = \frac{{\cal X}^\mu_{i+1,j}-{\cal X}^\mu_{i-1,j}}{2\Delta\tau}
\ee
\be
(\pa^2_{\tau}{\cal X}^\mu)_{i,j} = \frac{{\cal X}^\mu_{i+1,j}-2{\cal X}^\mu_{i,j}+{\cal X}
^\mu_{i-1,j}}{\Delta\tau^2}\ ,
\ee
where index $i$ marks the $i$-th time step and $j$ marks the $j$-th grid point in the sigma
direction.
Similar expressions are used for the spatial derivatives except for the points lying
at the boundary between dense and ``normal'' zones. In that case we must use modified formulas
\be
(\pa_{\sigma}{\cal X}^\mu)_{i,j} = \frac{1}{\Delta\sigma_{j-1}+\Delta\sigma_{j+1}}
\left(\frac{\Delta\sigma_{j-1}}{\Delta\sigma_{j+1}}({\cal X}^\mu_{i,j+1}-{\cal X}^\mu_{i,j})
+\frac{\Delta\sigma_{j+1}}{\Delta\sigma_{j-1}}({\cal X}^\mu_{i,j}-{\cal X}^\mu_{i,j-1})\right)
\ee
\be
(\pa^2_{\sigma}{\cal X}^\mu)_{i,j} = \frac{2}{\Delta\sigma_{j-1}+\Delta\sigma_{j+1}}
\left(\frac{1}{\Delta\sigma_{j+1}}({\cal X}^\mu_{i,j+1}-{\cal X}^\mu_{i,j})
-\frac{1}{\Delta\sigma_{j-1}}({\cal X}^\mu_{i,j}-{\cal X}^\mu_{i,j-1})\right)\ ,
\ee
where $\Delta\sigma_{j-1}=\sigma_j-\sigma_{j-1}$ and $\Delta\sigma_{j+1}=\sigma_{j+1}-\sigma_{j}$.

After the discretization process we obtain a set of four coupled non-linear equations for 
the four unknown variables ${\cal X}^\mu_{i+1,j}$ at each grid point $\sigma_j$. 
Since the equations for different $j$'s are independent of each other we can solve them in parallel.

\end{document}